\documentclass{PoS}
\usepackage{psfrag}
\usepackage{epsfig}
\newcommand{\Str}{{\rm Str}}
\newcommand{\Abar}{{\bar{A}}}
\newcommand{\Bbar}{{\bar{B}}}

\newcommand{\mpis}{{m_\pi^2}}
\newcommand{\metas}{{m_\eta^2}}
\newcommand{\mKs}{{m_K^2}}

\newcommand{\bi}{\begin{itemize}}
\newcommand{\ei}{\end{itemize}}
\title{Quark disconnected diagrams in chiral perturbation theory - the scalar form factor}

\ShortTitle{Quark disconnected contributions in chiral perturbation theory - the scalar form factor}

\author{\speaker{Andreas J\"uttner}\\%
	School of Physics and Astronomy\\
        University of Southampton\\
	Highfield, SO17 1AJ Southampton, UK\\
        E-mail: \email{juettner@soton.ac.uk}}

\abstract{Expressions for the Wick contractions contributing to the
scalar pion form-factor were computed  model-independently
in chiral perturbation theory at 
next-to-leading order. The results reveal correlations amongst the 
different contractions in terms of low-energy constants
and allow for extrapolating lattice data for individual Wick contractions.
The quark disconnected contribution to the real part of the form factor 
turns out to be suppressed with 
respect to the quark connected one. The corresponding contribution 
to the scalar radius has the same size as the connected contribution and
can therefore not be neglected.

}

\FullConference{The 30th International Symposium on Lattice Field Theory\\
		 June 24-29,  2012\\
		 Cairns, Australia}

\begin{document}

\section{Introduction}
Quark disconnected Wick contractions constitute a considerable numerical problem 
for lattice QCD. An estimate of the full quark-propagator is needed 
and the signal-to-noise ratio typically leaves a lot to be desired.
In current simulations in the iso-spin symmetric limit  disconnected 
contractions contribute to a number of 
phenomenologically relevant quantities ($g-2$, $\pi\pi$-scattering, 
hadronic $K$-decays, \dots). The situation becomes worse 
once also iso-spin breaking effects are included in the simulations
since  disconnected contributions will then contribute in a larger variety 
and to many more quantities.

Given the difficulties in computing quark disconnected contractions 
to a satisfactory precision
in lattice
simu\-lations one often neglects them, thereby introducing an unknown systematic
effect. Large efforts therefore go into devising dedicated
algorithms for improving the numerical evaluation of quark-disconnected 
diagrams (e.g.~\cite{Foley:2005ac,Bali:2009hu,Vera}). 
Here, an analytical and model-independent approach for computing individual
Wick contractions in chiral perturbation theory which was 
presented in~\cite{DellaMorte:2010aq} 
is applied to the case of the scalar 
pion form factor~\cite{Juttner:2011ur} for QCD with $N_f=2$ and $N_f=2+1$ 
dynamical flavours. At 
next-to-leading order (NLO) in the chiral expansion the contribution 
of the disconnected contraction to the form-factor is numerically small. 
Both the disconnected and the connected contribution however contribute 
with roughly the same magnitude to the scalar radius. The connected and 
disconnected contributions turn out to be correlated in terms of low-energy
constants of the chiral Lagrangian. The expressions derived here can be used to 
guide chiral extrapolations of lattice results for individual Wick
contractions.
As an aside the argument allowing to compute the quark-connected part with
partially twisted boundary conditions~\cite{Bedaque:2004kc,Sachrajda:2004mi} 
for an improved momentum resolution for the scalar form factor is presented.
\section{Technique}
The scalar form factor of the pion is defined as
\begin{equation}
 \langle \pi^i(p')|\bar u u+\bar d d|\pi^j(p)\rangle =\delta^{ij}F_{S,2}(t)\,,
\end{equation}
where $t=(p'-p)^2$ is the squared momentum transfer between the initial and
final pion and the sub-script on the r.h.s. identifies the scalar form factor
for $N_f=2$ flavour QCD.
In lattice QCD the matrix element on the l.h.s. is computed in terms of the
ground-state contribution to the Fourier transform of the 
Euclidean 3-pt. function
\begin{equation}
 \left< {O}^i(z)S(y){{O}^i}^\dagger(x)\right>\,,
\end{equation}
constructed of the interpolating operators
$O^i(x)=\bar \psi_2 \tau^i\gamma_5 \psi_2(x)$ and 
$S(x)=\bar \psi_2(x)          \psi_2(x)$.  The $\psi^T_2=(u,d)$ are  
$SU(2)$ flavour vectors of $u$- and $d$-quarks and 
the matrices $\tau_i=\sigma_i/2$ are proportional to the Pauli matrices.
Following~\cite{DellaMorte:2010aq} the two types of Wick contractions
contributing to this correlation function are 
\begin{equation}\label{eq:decomp_quarks}
\langle \bar u \gamma_5 d\, \bar d d\, \bar d \gamma_5 u\rangle=
\langle \bar u \gamma_5 v\, \bar v d\, \bar d \gamma_5 u\rangle+
\langle \bar u \gamma_5 d\, \bar v v\, \bar d \gamma_5 u\rangle\,.
\end{equation}
Note the additional valence quark $v$ with $m_u=m_d=m_v$ that was introduced 
in order to construct the decomposition on the r.h.s.. At this stage the 
connected contribution consists entirely of flavour off-diagonal currents
and the argument presented in~\cite{Boyle:2008yd,DellaMorte:2010aq} 
allows to compute at least this 
contribution using partially twisted boundary conditions, thus improving
the momentum resolution. Each individual term
on the r.h.s. represents an unphysical correlation function in the unphysical 
theory, QCD with an additional valence quark $v$. The
expression in chiral effective theory for the ground state contribution to the
l.h.s. has been derived many years ago in chiral perturbation theory at 
NLO~\cite{Gasser:1983yg,Gasser:1984ux} and later at 
NNLO~\cite{Bijnens:1998fm,Bijnens:2003xg} in both $SU(2)$ and $SU(3)$ chiral
perturbation theory. The effective theory frame work for the computation of 
the corresponding contributions to the r.h.s. is partially quenched
chiral perturbation theory (PQ$\chi$PT)~\cite{Bernard:1992mk,Sharpe:2000bc}
which is a frame-work allowing to vary the sea-quark and valence-quark
content independently. While the individual terms on the r.h.s. represent
unphysical matrix elements, their sum represents a physical process. It is
therefore conceivable to use different methods for the computation of the terms
on the r.h.s.. The connected contribution is conveniently computed in 
a lattice simulation. Where the disconnected contribution is difficult to
compute numerically the method advocated here can be applied. 
\section{PQ$\chi$PT for the scalar form factor}
According to~\cite{Bernard:1992mk,Sharpe:2000bc} the chiral effective theory
for $N_f=2$ and $N_f=2+1$ 
QCD with an additional valence quark $v$ is developed around the
graded flavour symmetry groups $SU(3|1)$ and $SU(4|1)$, respectively.
The leading order chiral Lagrangian 
is~\cite{Gasser:1983yg,Bernard:1992mk,Sharpe:2000bc,Gasser:1984gg,Bernard:1993sv}
\begin{equation}\label{eq:L2}
\mathcal{L}^{(2)}=\frac{F^2}{4}\Str\left\{
	\partial_\mu U \partial^\mu U^\dagger\right\}
	+\frac {F^2}4  \Str\left\{\chi U^\dagger + U\chi^\dagger \right\}\,,
\end{equation}
where $\chi=2 B (s+M)$. The mass matrix has the form 
$ M={\rm diag}\left(m_q,m_q,m_q,m_q\right)$ in $SU(3|1)$ and
$M={\rm diag}\left(m_q,m_q,m_s,m_q,m_q\right)$ in $SU(4|1)$
and we define the external scalar source as $s=2T^a s^a$ (with $T^a$ a 
generator of the flavour group).
The relevant counter terms can be derived from the Lagrangian,
\begin{equation}\label{eq:L4}
\begin{array}{rcl}
\mathcal{L}^{(4)}&=&
	L_4\Str\Big\{\partial_\mu U (\partial^\mu U)^\dagger\Big\}
		\Str\Big\{\chi U^\dagger + U \chi^\dagger\Big\}	
	+L_5\Str\Big\{\big(\partial_\mu U (\partial^\mu U)^\dagger\big)
		\big(\chi U^\dagger+U\chi^\dagger\big)\Big\}\\[2mm]
	&&+L_6\Str\Big\{\chi U^\dagger + U\chi^\dagger\Big\}^2
	+L_8\Str\Big\{U\chi^\dagger U\chi^\dagger + \chi U^\dagger \chi U^\dagger\Big\}^2\,.
\end{array}
\end{equation}
The remaining calculation proceeds as in standard chiral perturbation 
theory. 
The known result for the full form factor is reproduced,
\begin{eqnarray}\label{eq:results_ud}
F^{\rm F}_{S,2}(t)&=&
  2 B \Big\{1 +\frac 1{F^2}\Big(-\frac 1{2} \Abar_\mpis 
  	\hspace{13mm}+\,\Lambda^{\rm F}_2  
	\hspace{23mm}+\frac {(2 t - \;\,\mpis)}2 \Bbar(\mpis,t) 
	\Big)\Big\}\,,\\[3mm]
 F_{S,3}^{{\rm F}}(t)&=&
	2B\Big\{
	1+\frac{1}{F^2}\Big(-\frac 12 \Abar_\mpis+\frac 16 \Abar_\metas
	+\,\Lambda_3^{{\rm F}}
	+\frac{\mpis}{18}\Bbar(\metas,t)
	+\frac {(2t\,\,- {\mpis})}2\Bbar(\mpis,t)
	+\frac t4 \Bbar(\mKs,t)
	\Big)\Big\}\,.\nonumber
\end{eqnarray}
The expressions for $\bar A_{m^2}$ and $\bar B(m^2,t)$ are standard and can be
found in~\cite{Juttner:2011ur}. 
The new results are the individual expression for the connected 
($F^C_{S,N_f}$) and disconnected contribution ($F^C_{S,N_f}$):
{
	\small
\begin{eqnarray}\label{eq:results_ud}
F^{\rm C}_{S,2}(t) &=& 
  2 B \Big\{1 +\frac 1{F^2}\Big(-\frac 1{2} \Abar_\mpis 
  	\hspace{12mm}+\Lambda^{\rm C}_2
	+\frac  {(t - 2\mpis)}2 \Bbar(\mpis,t) 
	 \Big)\Big\}\,,\nonumber\\[3mm]
F^{\rm D}_{S,2}(t)&=& 
2 B \Big\{0 +\frac 1{F^2}\Big(\hspace{24mm}+ \Lambda^{\rm D}_2
+\frac {(t + \;\,\mpis)}2 \Bbar(\mpis,t) 
	 \Big)\Big\}\,,\\
 F_{S,3}^{{\rm C}}(t)&=&
	2B\Big\{
	1+\frac 1{F^2}\Big(-\frac 12 \Abar_\mpis+ \frac 16 \Abar_\metas
	+\Lambda_3^{{\rm C}}
	+\frac {(t- 2\mpis )}2\Bbar(\mpis,t)
	+\frac t4 \Bbar(\mKs ,t)
	+\frac {\mpis}3 \Bbar(\metas,\mpis,t)
	\Big)\Big\}\,,\nonumber\\[3mm]
 F_{S,3}^{{\rm D}}(t)&=&
	2B\Big\{
	0+\frac 1{F^2}\Big(\hspace{24mm}
	+\Lambda_3^{{\rm D}}
	+\frac{( t+\;\,\mpis  )}2 \Bbar(\mpis,t)
	+\frac \mpis{18} \Bbar(\metas,t)
	-\frac {\mpis}3  \Bbar(\metas,\mpis,t)
	\Big)\Big\}\,.\nonumber
\end{eqnarray}
}
Note that for both $N_f=2$ and $N_f=2+1$,
$F_S^F=F_S^C+F_S^D$, as expected. The $\Lambda_{N_f}^{C/D}$ contain combinations of the low energy constants (LECs),

\begin{equation}
\hspace{-1.6mm}\begin{array}{l@{\hspace{1mm}}c@{\hspace{1mm}}l@{\hspace{0mm}}l@{\hspace{0mm}}l@{\hspace{0mm}}l@{\hspace{0mm}}l@{\hspace{0mm}}l@{\hspace{0mm}}l@{\hspace{0mm}}}
 \Lambda_2^{{\rm F}}&=&
	 4\Big\{
	\mpis(-8 \tilde L_4^r-4 \tilde L_5^r&+16 \tilde L_6^r &+8\tilde L_8^r)
	&+t    (2  \tilde L_4^r+ \tilde L_5^r)
	&\Big\}\,,\\[3mm]
\Lambda_2^{{\rm C}}&=&
	 4\Big\{
	\mpis(-4 \tilde L_4^r-4 \tilde L_5^r&+8 \tilde L_6^r &+8\tilde L_8^r)
	&+t     \hspace{12.5mm}\tilde L_5^r
	&\Big\}\,,\\[3mm]
\Lambda_2^{{\rm D}}&=&
	 4\Big\{
	\mpis (-4 \tilde L_4^r &+ 8 \tilde L_6^r)&
	&+ t\;\,2      \tilde L_4^r
	&\Big\}\,,
\end{array}
\end{equation}
and 
\begin{equation}
\hspace{-0.6mm}\begin{array}{l@{\hspace{1mm}}c@{\hspace{1mm}}l@{\hspace{0mm}}l@{\hspace{0mm}}l@{\hspace{0mm}}l@{\hspace{0mm}}l@{\hspace{0mm}}l@{\hspace{0mm}}l@{\hspace{0mm}}}
 \Lambda_3^{{\rm F}}&=&
	4\Big\{
	\mpis(-6 L_4^r-4 L_5^r&+12 L_6^r &+8 L_8^r)
	&+\mKs (-4 L_4^r+8 L_6^r )
	&+t    (2  L_4^r+ L_5^r)
	&\Big\}\,,\\[3mm]
\Lambda_3^{{\rm C}}&=&
	 4\Big\{
	\mpis(-2 L_4^r-4 L_5^r&+4 L_6^r &+8L_8^r)
	&+\mKs (-4 L_4^r+8 L_6^r )
	&+t     \hspace{12.5mm}L_5^r
	&\Big\}\,,\\[3mm]
\Lambda_3^{{\rm D}}&=&
	 4\Big\{
	\mpis (-4 L_4^r &+ 8 L_6^r)&
	&&+ t\;\,2      L_4^r
	&\Big\}\,.
\end{array}
\end{equation}
The $L_i^r$ are the LECs of $SU(3)$ chiral perturbation theory~\cite{Gasser:1984gg} and the
$\tilde L_i^r$ are related to the better-known $SU(2)$ LECs $l_i^r$ via matching,
\begin{equation}
\begin{array}{l@{\hspace{1mm}}c@{\hspace{1mm}}l@{\hspace{0mm}}l@{\hspace{0mm}}l@{\hspace{0mm}}l@{\hspace{0mm}}l@{\hspace{0mm}}l@{\hspace{0mm}}l@{\hspace{0mm}}l@{\hspace{0mm}}}
\Lambda^{\rm F}_2&=
        &t\;\, l_4^r& &+ 4\mpis\;\, l_3^r\,, \\[3mm]
\Lambda^{\rm C}_2 &= 
        &t ( l_4^r&-8\tilde L_4^r)
	 &+ 4\mpis(l_3^r&+4\tilde L_4^r&-8\tilde L_6^r)\,,\\[3mm]
\Lambda^{\rm D}_2&= 
        &t (      &+8\tilde L_4^r)
	 &+ 4\mpis(       &-4\tilde L_4^r&+8\tilde L_6^r)\,.
\end{array}
\end{equation}
The remaining $SU(3|1)$ constants $\tilde L_4^r$ and $\tilde L_6^r$ 
are less well known since the 
corresponding terms in the $L^{(4)}$-Lagrangian can be removed in the
case of the flavour group $SU(2)$ (trace-identities).
For $N_f=2+1$ similar expressions for the individual Wick contractions
can be derived for the octet and singlet form 
form factors
\begin{equation}\label{eq:octet_singlet}
\begin{array}{lcl}
\langle\pi^i|\bar uu+\bar dd-2\bar ss|\pi^k\rangle&=&\delta^{ik}F_S^8(t)\,,\\[3mm]
\langle\pi^i|\bar uu+\bar dd+\bar ss|\pi^k\rangle &=&\delta^{ik}F_S^0(t)\,.\\[3mm]
\end{array}
\end{equation}

From the expressions in eq.~(\ref{eq:results_ud}) we see that while the
connected contribution to the form factor starts at leading order, the
disconnected contribution starts at NLO and is therefore expected to be smaller
in magnitude. 
Turning to the scalar radius 
$\langle r^2\rangle=6 \frac {dF_S(t)}{dt}|_{t=0}$, the derivative with
respect to the momentum removes the leading term from the connected
contribution and therefore 
connected and disconnected contribution start contributing
at the same order in the chiral power counting.

\begin{figure}
\begin{center}
 \psfrag{Epipisq}[t][t][1][0]{$t\,\rm [GeV^2]$}
 \psfrag{err}[c][t][1][0]{\hspace*{-32mm}\mbox{}\large
			$\Delta^{\rm Re}$}
 \psfrag{fpipi}[c][t][1][0]{\hspace*{0mm}\mbox{}\normalsize
			${\rm Re}\left(F_{S,3}^{\rm F,C,D}(t)/(2B)\right)$}
 \epsfig{scale=0.63,file=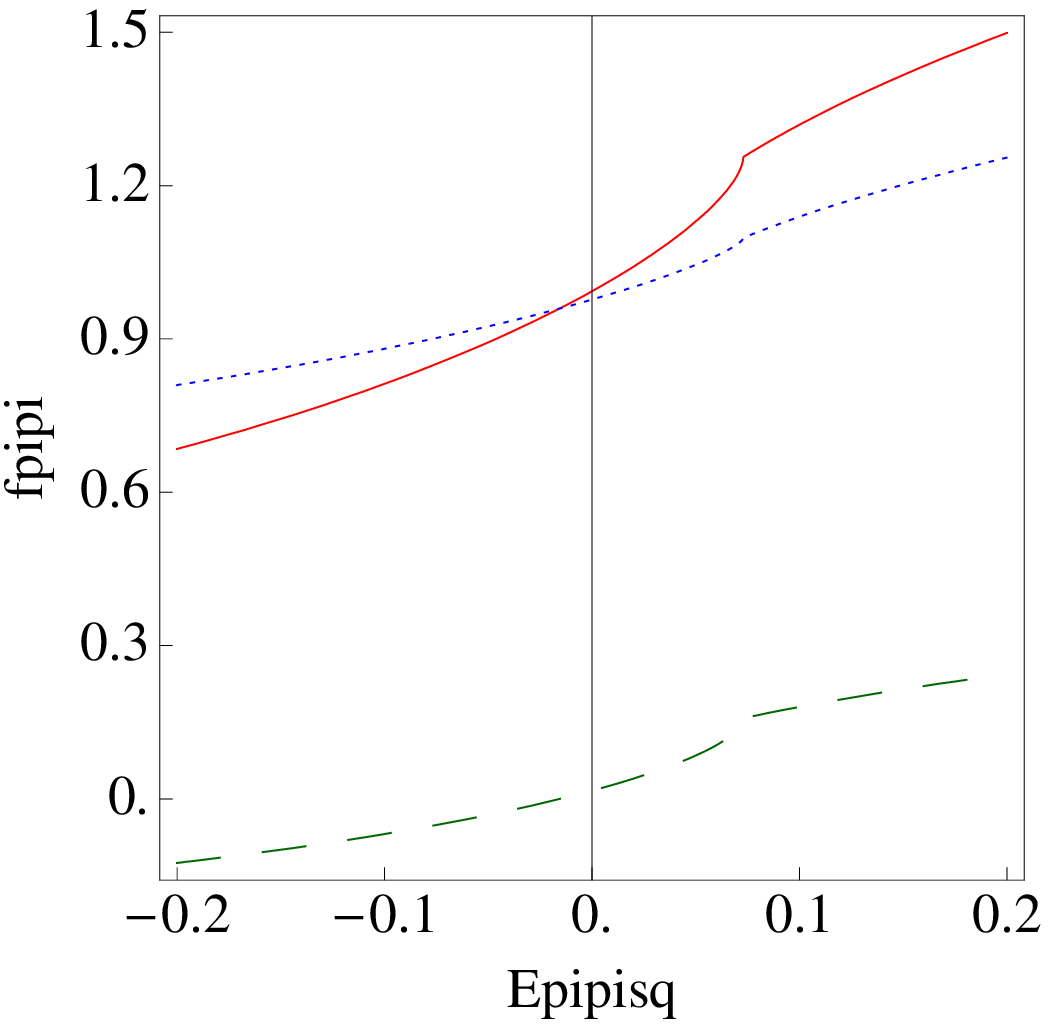}
 \psfrag{fpipi}[c][t][1][0]{\hspace*{0mm}\mbox{}\normalsize
			${\rm Im}\left(F_{S,2}^{\rm F,C,D}(t)/(2B)\right)$}
 \psfrag{fpipi}[c][t][1][0]{\hspace*{0mm}\mbox{}\normalsize
			${\rm Im}\left(F_{S,3}^{\rm F,C,D}(t)/(2B)\right)$}
		\hspace{5mm}\epsfig{scale=0.63,file=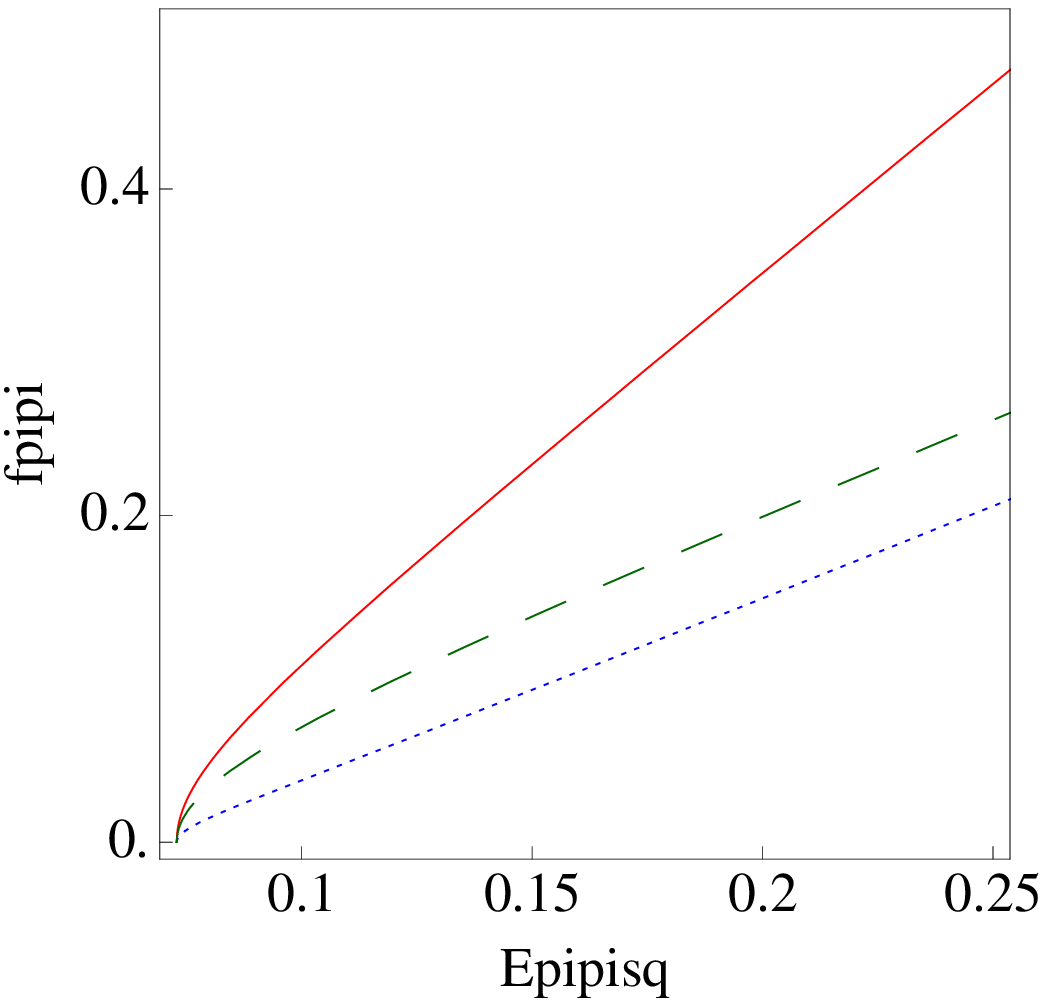}\\[6mm]
 \psfrag{rsqpi}[c][t][1][0]{\hspace{9mm}$\langle r^2\rangle \;[{\rm fm}^2]$}
 \psfrag{mpisq0}[t][t][1][0]{$m_\pi^2$ [GeV]$^2$}
 \epsfig{scale=0.63,file=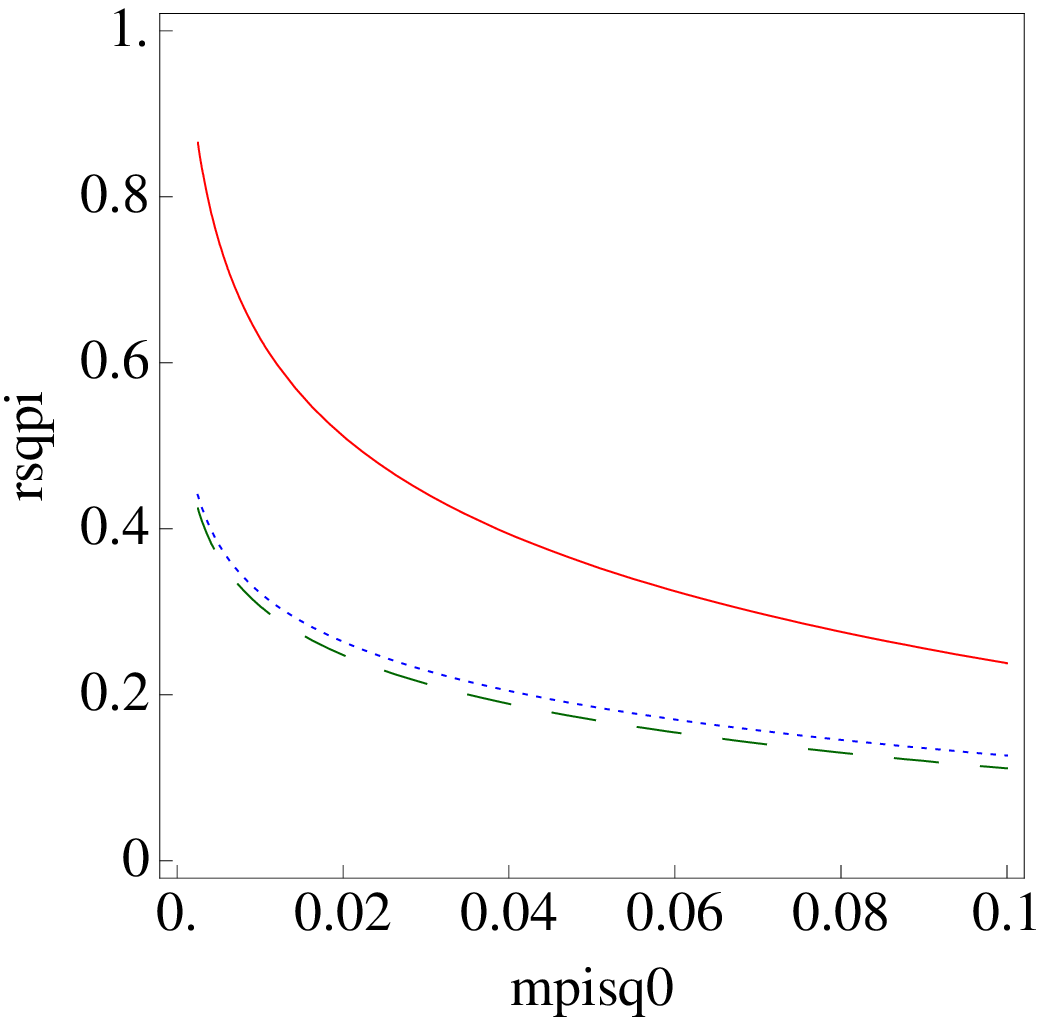}
\end{center}
\caption{First row: the plots show the momentum dependence of the real 
	(left) and imaginary (right)
	parts of the scalar form factor for 
	$N_f=2+1$: 
	Full form factor (solid red),
	connected contribution (dotted blue) and disconnected contribution 
	(dashed green).
	The bottom plot shows the scalar radius  for 
	$N_f=2+1$ as a function of the
	squared pion mass ($m_K$ is fixed to the physical value
	and $\metas=\frac 43 \mKs-\frac 13\mpis$, same color coding as first
	row). 
	}
\label{fig:results}
\end{figure}

Concentrating on the $N_f=2+1$-case we now
fix the free parameters in the expression for
$F_{S,3}^{F/C/D}(t)$ with input as summarised in table~\ref{tab:LECs}. 
The results for the 2-flavour case, which turn out to be 
very similar both qualitatively and quantitatively,
can be obtained after fixing the $\tilde L^r_i$ through matching 
the unphysical LECs  to the 3-flavour theory~\cite{Juttner:2011ur}.
\begin{table}
\begin{center}
\begin{tabular}{lllllll}
\hline\hline&&\\[-4mm]
$L_4^r$&\cite{Allton:2008pn,Colangelo:2010et}&$0.14\times10^{-3}$\\
$L_5^r$&\cite{Allton:2008pn,Colangelo:2010et}&$0.87\times10^{-3}$\\
$L_6^r$&\cite{Allton:2008pn,Colangelo:2010et}&$0.07\times10^{-3}$\\
$L_8^r$&\cite{Allton:2008pn,Colangelo:2010et}&$0.56\times10^{-3}$\\[1mm]
\hline
\hline
\end{tabular}
\caption{Values for the 3-flavour
low energy constants used in the illustration of 
the results. The subtraction scale is $\mu=0.77$GeV.}\label{tab:LECs}
\end{center}
\end{table}
The results are shown in the plots in figure~\ref{fig:results} for 
the full form factor and the connected and disconnected contribution, 
respectively.  For the real part the magnitude of the disconnected 
contribution is sub-dominant in line with the above observations. 
Above the two-pion threshold the disconnected contribution also contributes
to the imaginary part and is similar in magnitude to the connected contribution.
Since the leading contribution has no imaginary part this does not come as a 
surprise - 
the imaginary part in both the connected and the disconnected contribution
start at the same order, NLO.
We also show the mass-dependence of the scalar radius. What could have been
anticipated from the first of the three plots by observing that the slope around
$t=0$ of the connected and disconnected contribution is very similar,
is confirmed in the third plot - the connected and disconnected contributions
to the scalar radius are nearly of the same size. For a lattice simulation
this underlines that the disconnected contribution
cannot be neglected~\cite{Aoki:2009qn,Vera}.

\section{Conclusion}
Partially quenched chiral perturbation theory proves to be a 
powerful tool for understanding in detail the dynamics in the low-energy
sector of QCD. 
The decomposition of correlation functions into individual Wick contractions
provides guidance for lattice computations in various ways: First of all
an estimate of the magnitude of the disconnected contribution can be 
made in a model-independent way. Secondly, NLO LECs can be extracted from 
only the connected contribution to the form factor which is numerically
much better accessible in lattice simulations. The expression does however
contain new linear combinations of LECs and one has to study whether the 
LEC one is interested in can be extracted. 
Thirdly, a full computation of the form factor in lattice QCD is still very 
challenging due to the numerical cost of computing the disconnected
contribution to a satisfactory precision. 
In this situation one can either 
compute only the connected contribution in lattice QCD and predict the 
disconnected one, or one computes the connected contribution for a large set of 
parameters, while the disconnected one only for a reduced set of parameters
to fix the LECs and to eventually extrapolate it to the physical point.
\\[3mm]

\noindent{\bf Acknowledgements:}
The research leading to these results has received funding from the Euro\-pe\-an
Research Council under the European Union's Seventh Framework Programme
(FP7/2007-2013) / ERC Grant agreement 279757.
\bibliographystyle{JHEP}
\bibliography{proc}

\end{document}